\documentclass[conference,a4paper]{IEEEtran}

\usepackage{amssymb}
\usepackage{listings}
\usepackage{subfig}
\usepackage[utf8]{inputenc} 
\usepackage[T1]{fontenc}
\usepackage{url}              
\usepackage{cite}             

\usepackage[cmex10]{amsmath}  
\usepackage{mleftright}       
\mleftright                   
\usepackage{multirow}
\usepackage{multicol}
\usepackage{graphicx}         
\usepackage{booktabs}         

\DeclareMathOperator*{\argmin}{argmin}

\def\*#1{\mathbf{#1}}
\newcommand{\diag}{\mathrm{diag}}
\begin{document}

\title{On the Existence of Cyclic Lattice Codes} 

\author{%
  \IEEEauthorblockN{Chengpin Luo and Brian M. Kurkoski}
  \IEEEauthorblockA{%
    Japan Advanced Institute of Science and Technology\\
    923-1292 Nomi, Japan\\
    Email: \{luo.chengpin, kurkoski\}@jaist.ac.jp}
}

\maketitle

\begin{abstract}
A coding lattice $\Lambda_c$ and a shaping lattice $\Lambda_s$ forms a nested lattice code $\mathcal{C}$ if $\Lambda_s \subseteq \Lambda_c$. Under some conditions, $\mathcal{C}$ is a finite cyclic group formed by rectangular encoding. This paper presents the conditions for the existence of such $\mathcal{C}$ and provides some designs. These designs correspond to solutions to linear Diophantine equations so that a cyclic lattice code $\mathcal C$ of arbitrary codebook size $M$ can possess group isomorphism, which is an essential property for a nested lattice code to be applied in physical layer network relaying techniques such as compute and forward.
\end{abstract}

\section{Introduction}
\label{sec:double-blind-policy}

For two $n$-dimensional lattices ${\Lambda_c}$ and ${\Lambda_s}$, which are discrete additive subgroups of $\mathbb{R}^n$, a quotient group $\Lambda_c / \Lambda_s$ is formed if $\Lambda_s \subseteq \Lambda_c$. Denote the coset leaders from the zero-centered Voronoi region of $\Lambda_s$ as $\mathcal{C}$; $\mathcal{C}$ is also called the nested lattice code. A suitable choice of $\Lambda_c$ and $\Lambda_s$ provides coding gain and shaping gain respectively for wireless communication systems. $\mathcal{C}$ is also a candidate for physical-layer networking coding due to its group properties. 

The encoding of information is to map integers to a codeword of $\mathcal{C}$. The inverse operation of encoding is called indexing. While the quotient group always exists when $\Lambda_c \subseteq \Lambda_s$,  efficient encoding may exist only under certain conditions.  Conway and Sloane provided an efficient way to perform encoding and indexing for self-similar lattices, that is, $\Lambda_s = K \Lambda_c$ with $K$ a non-negative integer \cite{1056761}. However, there may be practical reasons to not use self-similar lattices and rectangular encoding generalizes this encoding and indexing to lattices which are not self-similar \cite{8361838}.  One special case is when the nested lattice code forms a cyclic group under rectangular encoding, which we call a cyclic lattice code in this paper. One example of a cyclic lattice code was given in \cite{8361838}. However, it is still not clear under what conditions a cyclic lattice code exists. In this paper, we derive such conditions. 

Furthermore, a lattice code with group isomorphism is required by compute-and-forward relaying for physical layer network coding \cite{6034734}. An additional contribution of this paper is conditions for a cyclic lattice code to possess group isomorphism, which allows construction of $\mathcal{C}$ of arbitrary codebook size.

One benefit of cyclic lattice codes is a reduced penalty due to mapping bits to $\mathcal C$, when $|\mathcal C|$ is not a power of two. Cyclic lattice codes also admit a simplified encoder structure, requiring a single generator vector, rather than a generator matrix.  

The remainder of this paper is organized as follows. Section II gives preliminaries including definitions of nested lattice codes, rectangular encoding and cyclic lattice codes. Section III derives the conditions for the existence of cyclic lattices, based on which one design for $n=2$ cyclic lattice code as well as two designs for general $n$ are provided. Section IV studies the isomorphism of a cyclic lattice code, illustrated by a few examples. Section V concludes the paper.

\section{Preliminaries}
\label{sec:page-limit}

\subsection{Lattices}
\label{sec:difinition lattice matrix}
\label{sec:fundamental region}

An $n$-dimension point $\*y$ of lattice $\Lambda$ is formed by $\*y = \*G\*b$ where $\*G \in \mathbb{R}^{n \times n}$ is a full rank generator matrix and $\*b \in \mathbb{Z}^{n}$ is an integer vector. The check matrix is defined as $\*H = \*G^{-1}$, so that $\*{Hy = b}$ holds. For two lattices $\Lambda_c$ and $\Lambda_s$, their generator matrices are respectively denoted as $\*G_c$ and $\*G_s$. Correspondingly, their check matrices are denoted as $\*H_c$ and $\*H_s$. The sublattice condition $\Lambda_s \subseteq \Lambda_c$ is satisfied if and only if $\*H_c\*G_s$ is a matrix of integers \cite{zamir_nazer_kochman_bistritz_2014}.

A region $\mathcal{F}$ is a fundamental region of lattice $\Lambda$ if shifts of $\mathcal{F}$ by all $\*y \in \Lambda$ exactly covers the whole $\mathbb{R}^n$. One example is the Voronoi region $\mathcal{V}$. Another example is a parallelotope $\mathcal{P}(\*P)$ which has the following form:
\begin{equation}\label{eqn:parallelotope}
 \mathcal{P}(\*P) = \left \{ s_1\*p_1 + s_2\*p_2 + \dots + s_n\*p_n | 0 \leq s_i <  1 \right \}   
\end{equation}
for some matrix $\*P = 
\begin{bmatrix}
 \*p_1 & \*p_2 & \dots & \*p_n
\end{bmatrix}$. Clearly, $\mathcal{P}(\*G)$ is a fundamental region for lattice $\Lambda$ with generator matrix $\*G$. All fundamental regions have same volume $|\text{det}(\*G)|$.

\subsection{Nested Lattice Codes}
\label{sec:neseted lattice codes}  
Assuming $\Lambda_s \subseteq \Lambda_c$, any point $\*y \in \Lambda_c$ can be used to form a set $\*y + \Lambda_s$, which is called a coset of $\Lambda_s$ in $\Lambda_c$ \cite{21245}. A quotient group $\Lambda_c / \Lambda_s$ is then defined as the set of cosets:
\begin{equation}
    \Lambda_c / \Lambda_s = \left \{ \*y + \Lambda_s | \*y \in \Lambda_c\right \}.
\end{equation}
A coset leader is a representative selected from one coset. The intersection of $\Lambda_c$ and any $\mathcal{F}$ of $\Lambda_s$ forms the set of coset leaders. Choosing $\mathcal{F}$ to be the zero-centered Voronoi region $\mathcal{V}$, then the coset leaders of $\Lambda_c / \Lambda_s$ become codewords used in a communication system:
\begin{equation}
    \mathcal{C} = \Lambda_c \cap \mathcal{V}
\end{equation}
which is called a nested lattice code. The size of $\mathcal{C}$ is $M = |\text{det}(\*G_s)| / |\text{det}(\*G_c)|$. The code rate is therefore:
\begin{equation}
    R = \frac{1}{n} \log_2 \frac{|\text{det}(\*G_s)|}{|\text{det}(\*G_c)|}.
\end{equation}

\subsection{Rectangular Encoding}
\label{sec:retangular encoding}
The nested lattice code $\mathcal{C}$ has a rectangular encoding if there exists an encoding matrix $\*M = \diag(M_1, M_2,...,M_n)$ such that  an integer vector $\*b = [b_1, b_2, ..., b_n]^T$ with each $b_i \in \left \{ 0, 1, ..., M_i - 1\right \}$ can be bijectively mapped to codewords $\*y \in \mathcal{C}$ by 
\begin{equation}
    \label{eqn:lattice mod}
    \*y = \*G_c\*b - Q_{\Lambda_s}(\*G_c\*b),
\end{equation}
where $Q_{\Lambda_s}$ is the quantization function:
\begin{equation}
    Q_{\Lambda_s}(\*y) = \argmin_{\*x \in \Lambda_s} \left \| \*y - \*x  \right \|^2.  
\end{equation} 
 $M_i-1$ is the maximum integer that can be encoded in coordinate $i$ and the codebook size is $\left | \mathcal{C} \right | = M = \prod_{i=1}^{n}M_i$. Equation (\ref{eqn:lattice mod}) can also be written as $\*x \bmod \Lambda_s$, where $\*x = \*G_c\*b$. Two tests for lattice membership are used; the following three statements are equivalent: (1) $\*x \in \Lambda$ (2) $\*H \*x$ is a vector of integers (3) $\*x \bmod \Lambda = \*0$.

From (\ref{eqn:parallelotope}) and (\ref{eqn:lattice mod}), it is not hard to see that the point $\*G_c\*b \in \Lambda_c$ is inside the parallelotope $\mathcal{P}(\*G_c\*M)$ \cite{8361838}. To have a bijective mapping between $\mathcal{P}(\*G_c\*M)$ and $\mathcal{C}$, one sufficient condition is to make $\mathcal{P}(\*G_c\*M)$ a fundamental region of $\Lambda_s$ \cite[Lemma 4]{8361838}. Discussions about designing $\*M$ for triangular matrix lattices or full matrix lattices were also given in \cite{8361838}.

\subsection{Cyclic Groups, Cyclic Lattice Codes}
A cyclic group is a group which can be generated by a single element $\*g$. Such an element $\*g$ is called a generator \cite{alonso1991notes}. The order of a cyclic group is the number of elements it contains. Elements of a finite cyclic group of order $M$ can be represented by integer multiples of $\*g$. The generator of a cyclic group is not necessarily unique. If $\*g$ is a generator, any $k  \*g$ is a generator for $k \in \left \{0,1,..., M-1\right\}$ and $\gcd(k, M) = 1$, where $\gcd(\cdot)$ is the greatest common divisor of two or more integers. For a vector $\*s$, $\gcd(\*s)$ is interpreted as the $\gcd$ of the elements of $\*s$.

A nested lattice code $\mathcal{C}$ is called a cyclic lattice code\footnote{In this paper, the term ``cyclic lattice code'' refers to a cyclic group and in general they are not cyclic codes which is a block code where a cyclic shift of each codeword also gives a codeword. Micciancio studied cyclic lattices which are defined analogously to cyclic codes \cite{Micciancio-FOCS02}.} in coordinate $k$ if it is a cyclic group of order $M$ and can be generated by $\*g_k$, which is the column $k$ of $\*G_c$. In other words, the whole codebook $\mathcal{C}$ can be generated from information vector $\*b = \begin{bmatrix}
0 & \dots & 0 & b_k & 0 & \dots & 0
\end{bmatrix}^T$ where $b_k \in \left \{0,1,...,M-1\right \}$ with operation 
\begin{equation}
    \*y = b_k \*g_k - Q_{\Lambda_s}(b_k\*g_k)
\end{equation}
where $\*y \in \mathcal{C}$. In this case, the encoding matrix $\*M$ satisfies $M_k = M$ and $M_i = 1$ for any $i \neq k$ with  $i,k \in \{1,\dots, n \}$ \cite{8361838}. 


\subsection{Motivation: Efficient Encoding of Bits}

Mapping bits to codewords is a fundamental aspect of any communication system.  Nominal compute-forward systems use $\Lambda / p \Lambda$, where $p$ is a prime number. While $\log_2(p)$ bits are encoded per coordinate for self-similar lattices, when $p$ is prime, there may be a significant loss as not all codewords can be efficiently mapped using binary information. 

One motivation of using cyclic lattice code is to improve the codeword usage compared with the non-cyclic case such as self-similar lattice codes. The codeword usage for cyclic lattice code is: 
\begin{equation}
    U_c = \frac{2 ^ {\left \lfloor \log_2(M) \right \rfloor}}{M}.
\end{equation}
A self-similar nested lattice is defined as $\Lambda_c / K \Lambda_c$, for which $M_i = K$ for all $i$. In each coordinate, up to $\left \lfloor \log_2(M_i) \right \rfloor$ bits of information could be encoded. Thus, the total codeword usage for self-similar nested lattice is:
\begin{equation}
    U_s = \frac{2 ^ {n \left \lfloor \log_2(M_i) \right \rfloor}}{M}. 
\end{equation}
Since $\left \lfloor \log_2(M) \right \rfloor = \left \lfloor n \log_2(M_i)\right \rfloor \geq n \left \lfloor \log_2(M_i)\right \rfloor$, $U_c \geq U_s$. One example of such improvement could be $n=2, K=7$,  $U_c - U_s = 32.65\%$.

\section{Existence of Cyclic Lattices}
\label{sec:paper-format}

Lemma 1 is a preliminary step needed for proving the central result Proposition 2. 

\textit{Lemma 1}: Consider an $n$-dimensional lattice $\Lambda_s$ with generator matrix $\*G_s = \begin{bmatrix}
\*g_1 & \*g_2 & \dots & \*g_n
\end{bmatrix}$.  The line segment with endpoint $\*0$ and $\*y = \*G_s \cdot \*b$ with $\*b = \begin{bmatrix}
b_1 & b_2 & \dots & b_n
\end{bmatrix}^T$ does not intersect any other point of $\Lambda_s$ if and only if $\gcd(b_1, b_2, ..., b_n) = 1$. 

\textit{Proof:} 
The \textit{only if} direction: If $\gcd(b_1, b_2, ..., b_n) \geq k$ where $k > 1$, then $\*y = k\*y'$, where $\*y$ are indexed by $(\frac{b_1}{k}, \frac{b_2}{k},...,\frac{b_n}{k})$. Therefore, the line segment between $\*y$ and $(0,0)$ must intersect lattice point $\*y'$.  The \textit{if} direction: Assume $\gcd(b_1, b_2,\cdots,b_n) = 1$, and assume there is a lattice point of $\Lambda_s$ in between, which is given by $\*x = \*G_s \cdot 
\begin{bmatrix}
\frac{c}{d}\cdot b_1 & \frac{c}{d}\cdot b_2 & \dots &   \frac{c}{d}\cdot b_n
\end{bmatrix}^T$ where $c, d$ are coprime integers and $\frac{c}{d} \in (0,1)$. Since $\frac{cb_1}{d}, \cdots, \frac{cb_n}{d}$ are all integers, $\gcd(b_1,...,b_n) \geq d$ which contradicts our initial assumption. Therefore, there must be no lattice points between $\*0$ and $\*b$.

Consider a cyclic lattice code $\mathcal{C}$ with $\Lambda_s \subseteq \Lambda_c$. Without loss of generality, we choose $\*g_n$, column $n$ of $\*G_c$, as the generator of the cyclic group. Because the group generator is $\*g_n$, $M \cdot \*g_n \bmod  \Lambda_s = \*0$ holds, and so it must be that
\begin{equation}
\label{eqn:existproof1}
    \*H_s\*G_c \cdot  \begin{bmatrix}
0 & \dots & 0& M
\end{bmatrix}^T = 
 \begin{bmatrix}
s_1 & \dots & s_n 
\end{bmatrix}^T
\end{equation}
are integers; we define $s_1,\ldots,s_n$ to be these integers.


$\mathcal{P}(\*G_c\*M)$ is a fundamental region of $\Lambda_s$ if and only if $ k \cdot \*g_n \bmod  \Lambda_s \neq \*0$ for $k=1,\dots, M-1$. Following Lemma 1, we could therefore see that $\mathcal{C}$ is cyclic if and only if $\gcd(s_1,\ldots,s_n) = 1$. Since $\*H_s\*G_c = (\*H_c\*G_s)^{-1}$, (\ref{eqn:existproof1}) could be written as 
\begin{equation}
\label{eqn:existproofn1}
    \frac{1}{\text{det}(\*H_c\*G_s)} \begin{bmatrix}
 \*q_1 & \dots & \*q_n
\end{bmatrix} \begin{bmatrix}
0 \\
\vdots \\
0 \\
M\\

\end{bmatrix} = 
 \begin{bmatrix}
s_1 \\ 
\vdots \\ 
s_n 
\end{bmatrix},
\end{equation}
where $\*q_i$ is column $i$ of $\text{det}(\*H_c\*G_s)(\*H_c\*G_s)^{-1}$. Since $\text{det}(\*H_c\*G_s) = M$, we have 
\begin{equation}
\label{eqn:existnd}
    \*q_n =  \begin{bmatrix}
s_1 & 
\dots & 
s_n 
\end{bmatrix}^T.
\end{equation}

Finally, we could come to our conclusion on the existence of cyclic lattice code.

\textit{Proposition 2:} An $n$ dimensional nested lattice code $\mathcal{C}$ with $\Lambda_s \subseteq \Lambda_c$ is a cyclic lattice code in coordinate $i \in \left \{ 1,2,...,n\right \}$ if and only if $\gcd(\*q_i) = 1$, where $\*q_i$ is column $i$ of $\text{det}(\*H_c\*G_s)(\*H_c\*G_s)^{-1}$.

\subsection{$n=2$ Nested Lattice}
First, we would like to design an $n=2$ cyclic lattice code using Proposition 2. Suppose we have
\begin{equation}
    \*H_c\*G_s = \begin{bmatrix}
 a & b\\
 c &d
\end{bmatrix},
\end{equation} and $\*g_2$ of $\*G_c$ is chosen as the generator of the cyclic group, then (\ref{eqn:existnd}) becomes
\begin{equation}
    \begin{bmatrix}
-b & a
\end{bmatrix} =     \begin{bmatrix}
s_1 & s_2
\end{bmatrix}. 
\end{equation}

Specifically, the existence condition for $n=2$ cyclic lattice code could be described in the following lemma.

\textit{Lemma 3:} An $n=2$ lattice code $\mathcal{C}$ with $\Lambda_s \subseteq \Lambda_c$ is a cyclic lattice code in coordinate $i$ if and only if the elements in row $j$ of $\*H_c\*G_s$ are coprime, with $i \neq j$ and $i,j \in \{1,2\}$.

An example is given as follows. Consider $\Lambda_c$ has generator matrix:
\begin{equation}
    \label{eqn:Hc}
    \renewcommand\arraystretch{1.2}
    \*G_c = \begin{bmatrix}
 \frac{4}{3} & \frac{2}{9}\\
\frac{4}{3} &\frac{8}{9}
\end{bmatrix},
\end{equation}
and $\*H_c\*G_s$ has all rows with coprime entries:
\begin{equation}
    \label{eqn:example1}
    \*H_c\*G_s = \begin{bmatrix}
 4 & 9\\
3 &  8
\end{bmatrix}.
\end{equation}

\begin{figure*}[htbp]
  \centering
  \subfloat[cyclic lattice encoding with $\*M = \diag(1, 5)$]{%
    \label{fig:subfigA}
    \includegraphics[scale=0.5]{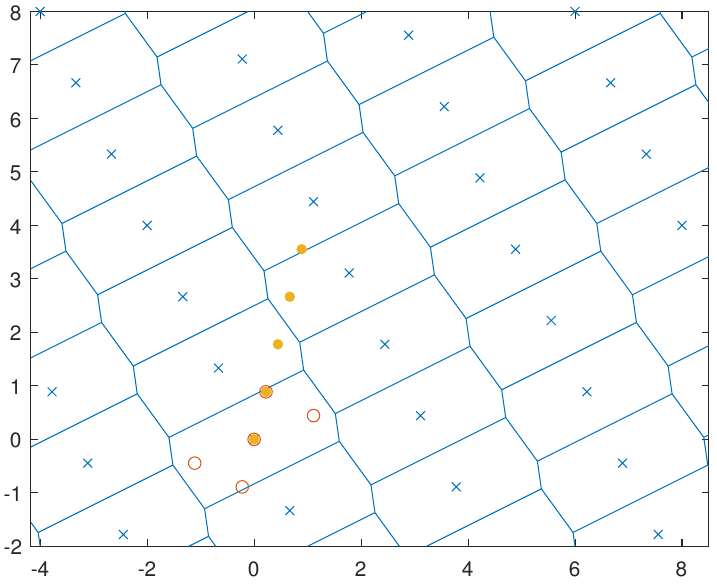}}
  \hfil
  \subfloat[cyclic lattice encoding with $\*M = \diag(5, 1)$]{%
    \label{fig:subfigB}
    \includegraphics[scale=0.5]{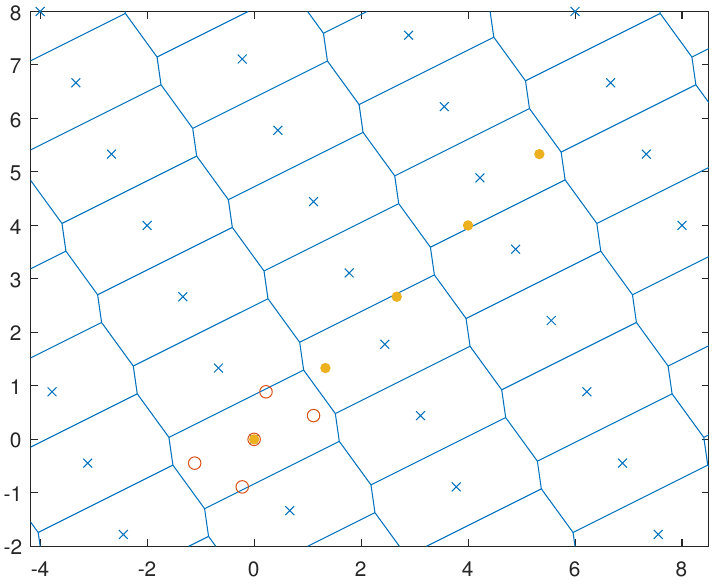}}  
  \caption{Encoding of cyclic lattice codes with $M=5$; Red circles: codewords $\mathcal{C}$; Yellow filled dots: lattice points $\*y_c =\*G_c\*b$ inside parallelotope $\mathcal{P}(\*G_c\*M)$; Blue cross: lattice points $\*y_s \in \Lambda_s$.}
  \label{fig:subfigures}
\end{figure*}

Since both pairs $(4, 9)$ and $(3, 8)$ are both relatively prime,  either coordinate $2$ or $1$ can be chosen respectively, i.e., $\*g_2 =\begin{bmatrix}
\frac{2}{9} & \frac{8}{9}
\end{bmatrix}^T$ or $\*g_1 =\begin{bmatrix}
\frac{4}{3} & \frac{4}{3}
\end{bmatrix}^T$ can be generators. The size of $\mathcal{C}$ is $M = 5$ and therefore we could have either $\*M = \diag(1, 5)$ for coordinate $2$ or $\*M = \diag(5, 1)$ for coordinate $1$. Applying (\ref{eqn:lattice mod}), we can obtain  the codebook $\mathcal{C}$, which are shown in Fig.~\ref{fig:subfigures}.

\subsection{Design for General $n$}
For general $n \geq 3$, there are more degrees of freedom to design $\*H_c\*G_s$ using Proposition 2. We would like to restrict $\*H_c\*G_s$ to make the design easier. Denote $\*W = \*H_c\*G_s$ and
 \begin{equation}
    \label{eqn:nd-isomorphic0}
     \*W = \begin{bmatrix}
 w_{1,1} & w_{1,2} & \dots & w_{1,n}\\
 w_{2,1} & w_{2,2} & \dots & w_{2,n}\\
 \vdots &  \vdots &  &\vdots \\
  w_{n-1,1} & w_{n-1,2} & \dots & w_{n-1,n}\\
w_{n,1} & w_{n,2} & \dots & w_{n,n}\\
\end{bmatrix}. 
 \end{equation}
Denote $\*W^{(j,k)}$ as the $n-1 \times n-1$ submatrix obtained by removing row $j$ and column $k$ of $\*W$, then $\text{det}(\*W^{(j,k)})$ is the minor and $C^{(j,k)} = (-1)^{(j+k)}\text{det}(\*W^{(j,k)})$ is the cofactor. Then, (\ref{eqn:existnd}) can be written as: 
 \begin{equation}
    \label{eqn:nd-isomorphic1}
    \begin{bmatrix}
 C^{(n,1)}\\
 \vdots \\
 C^{(n,n-1)}\\
C^{(n,n)}\\
\end{bmatrix} = 
 \begin{bmatrix}
s_1 \\ 
\vdots \\ 
s_n
\end{bmatrix}.
 \end{equation}
 A sufficient condition to have $\gcd(s_1, s_2, ...,s_n) =1$ is that any subset $\mathcal S' \subseteq \left \{s_1, s_2, ...,s_n  \right \}$ has $\gcd(\mathcal S') = 1$, where $ 2 \leq |\mathcal S'| \leq n$.  We give simple designs of the $\*W$ matrix for $|\mathcal S'| = 2$ and $|\mathcal S'|=3$.
 
 \subsubsection{$|\mathcal S'| = 2$}
 Assume $\mathcal S'=\{s_{n-1},s_n \}$ and $s_{n-1}, s_n$ are coprime. Since $\*W^{(n, n-1)}$ and $\*W^{(n, n)}$ only differ in the last column, we design $\*W$ with the following structure:
 \begin{equation}
    \label{eqn:ndexist1}
     \*W = \begin{bmatrix}
 0 &  \dots & 0 & a & b\\
 1 & \dots & 0 & w_{2,n-1} & w_{2,n}\\ 
 \vdots &  \ddots & \vdots  &\vdots & \vdots\\
  0 & \dots & 1 & w_{n-1,n-1} & w_{n-1,n}\\
w_{n,1} & \dots & w_{n,n-2} & w_{n,n-1} & w_{n,n}\\
\end{bmatrix},
 \end{equation}
 where $\gcd(a,b) = 1$. The first row except $a,b$ are all zeros, the submatrix from row $2$ to row $n-1$, column $1$ to column $n-2$ is a unimodular matrix which we simply choose as the identity matrix, the remaining values are free to choose as long as $\*W$ remains to be full rank. With this special structure, it is trivial to compute 
$C^{(n,n-1)} = -\text{det}(\*W^{(n, n-1)}) = -b$ and $C^{(n,n)}=\text{det}(\*W^{(n, n)}) = a$. Because $a, b$ are relatively prime, $\gcd(C^{(n,n-1)}, C^{(n,n)}) = 1$, satisfying the conditions of Proposition 2.

\subsubsection{$|\mathcal S'| = 3$}
Assume $\mathcal S'=\{s_{n-2}, s_{n-1},s_n \}$ and $\*W$ satisfy the following structure:
\begin{equation}
    \label{eqn:ndexist-fix3}
     \*W = \begin{bmatrix}
 0 &  \dots & 0 & a & b & c\\
 1 & \dots & 0 & 0 & 0 &0\\
 \vdots &  \ddots & \vdots  &\vdots & \vdots &\vdots\\
 0 & \dots & 1 & 0 & 0  &0\\
  0 & \dots & 0 & 0 & 1  &1\\
w_{n,1} & \dots & w_{n,n-3} & w_{n,n-2} & w_{n,n-1} & w_{n,n}\\
\end{bmatrix}.
 \end{equation}
 The three cofactors are then
 \begin{equation}
     (C^{(n,n-2)}, C^{(n,n-1)}, C^{(n,n)}) = (-b+c, a, -a).
 \end{equation}
Thus, if $\gcd(-b+c, a) = 1$ then Proposition 2 holds.

\enlargethispage{-1.4cm} 

\section{Conditions for Isomorphism Existence}
\label{sec:isomorphism}
Nazar and Gastpar \cite{6034734} proposed compute-and-forward (CF) techniques for physical-layer network coding. Several algebraic approaches have been proposed for constructing nested lattice codes which the CF framework relies on, e.g., \cite{5513739, tunali2015lattices, 7282518, 8387793,6952389}. In general, the group properties of nested lattice codes are essential. Linear labeling of $\Lambda_c$ with group linearity properties was proposed by Feng \emph{et al.} \cite{5513739}.  They also gave conditions on the existence of an isomorphism. In addition, there are conditions for group isomorphism based on the structure of the matrix $\*H_c\*G_s$ \cite{8361838}, which is given in Lemma 3 below.

Define group $\mathbb{Z}_{*}^{n}$ as:
\begin{equation}
    \mathbb{Z}_{*}^{n} = \mathbb{Z}_{M_1} \times \mathbb{Z}_{M_2} \times \dots \times \mathbb{Z}_{M_n},
\end{equation}
where $\mathbb{Z}_{M_i} = \mathbb{Z} / M_i \mathbb{Z} = \left \{0, 1, ..., M_i - 1 \right \}$.  
Consider two information vectors $\*b_1, \*b_2 \in \mathbb{Z}_{*}^{n}$. A group isomorphism is satisfied if 
\begin{equation}
    \text{enc}(\*b_1 \boxplus \*b_2) = \text{enc}(\*b_1) \oplus \text{enc}(\*b_2), 
 \end{equation}
 where $\text{enc}(\cdot)$ is the same operation as in (\ref{eqn:lattice mod}), say, $\*y = \text{enc}(\*b)$; the operation $\boxplus$ is group addition performed coordinate-wise, i.e., for coordinate i, we have $b_{i,1} \oplus b_{i, 2} = (b_{i,1} \oplus b_{i, 2}) \text{  mod  }  {M_i}$; $\oplus$ is the group operation over $\mathcal{C}$.

 \textit{Lemma 3 \cite{8361838}: } For arbitrary nested lattice $\Lambda_s \subseteq \Lambda_c$, if all elements from row $i$ of $\*H_c\*G_s$ could be divided by $M_i$ for all $i = 1, 2,..., n$, then an isomorphism exists between group $\mathbb{Z}_{*}^{n}$ and $\mathcal{C}$ under group operation $\oplus$. 

 Clearly, $\*H_c\*G_s$ in (\ref{eqn:example1}) does not satisfy the condition in Lemma 3 under cyclic encodings since neither row is divisible by  $M = 5$. On the other hand, a self-similar lattice code $\Lambda / K\Lambda$ which could be used for compute-and-forward relaying \cite{sakzad2014phase} satisfies the isomorphism since its rectangular coding matrix $\*M = K\*I_n$ but it is not cyclic. 

 Assume a cyclic lattice code with the generator in coordinate $n$. According to Lemma 3, the last row of $\*H_c\*G_s$ should be divisible by $M$. We can write last row of $\*W = \*H_c\*G_s$ as
 \begin{equation}
    \label{eqn:nd-isomorphic}
     \*W_n = \begin{bmatrix}
r_1 M & \dots & r_iM & \dots & r_nM \\
\end{bmatrix}. 
 \end{equation}
 Since $\text{det}(\*W) = M$, we have the following constraint:
 \begin{equation}
    \label{eqn:isoconstr}
     \frac{1}{M}\text{det}(\*W)  = \sum_{i=1}^{n}r_iC^{(n,i)} = 1.
 \end{equation}
 
Eqn.~(\ref{eqn:isoconstr}) is a linear Diophantine equation in the variables $r_1, r_2, ..., r_n$, which might have no solutions or multiple solutions. For $\*W$ in (\ref{eqn:ndexist1}), setting $r_{n-1}, r_n \neq 0$ and the other $r_i=0$ gives rise to the linear Diophantine equation 
 \begin{equation}
     \label{eqn:isomorphismnd}
     -br_{n-1} + ar_n - 1 = 0.
 \end{equation} 
 Similarily, for $\*W$ in (\ref{eqn:ndexist-fix3}), keeping only the last three $r_i$ non-zero gives
  \begin{equation}
     \label{eqn:isomorphismnd-fix3}
     (-b+c)r_{n-2} + ar_{n-1} - ar_{n} = 0,
 \end{equation}
(\ref{eqn:isomorphismnd}) and (\ref{eqn:isomorphismnd-fix3}) are naturally solvable as long as the cyclic conditions are satisfied, i.e., $\gcd(a,b)=1$ and $\gcd(-b+c, a)=1$ holds, respectively. 
 Hence, we can come to Lemma 4.
 
 $\textit{Lemma 4:}$
 For a cyclic lattice code of arbitrary size $M$ with structure $\*W$ in (\ref{eqn:ndexist1}) or (\ref{eqn:ndexist-fix3}), there always exists an isomorphsim that can be obtained by  modifying the last row of $\*W$ to the form of (\ref{eqn:nd-isomorphic}), which is equivalent to finding a solution to (\ref{eqn:isomorphismnd}) and (\ref{eqn:isomorphismnd-fix3}) respectively and setting any other irrelevant $r_i$ to $0$.

Examples for constructing isomorphic cyclic lattice codes are given as follows.

\subsection{$n=2$ nested lattice}
 For the case of $n=2$,  $\*H_c\*G_s$ only has the last two columns of  (\ref{eqn:ndexist1}). Consider $\*M = \diag(1, M)$ and
 \begin{equation}
\label{eqn:isomorphic form}
\*H_c\*G_s = \begin{bmatrix}
 a & b\\
r_1 M & r_2 M \\
\end{bmatrix}
\end{equation}
 where $a$ and $b$ are coprime and both $r_1, r_2$ are integers. 
 
 

\begin{figure}[htbp]
  \centering
  \includegraphics[scale=0.5]{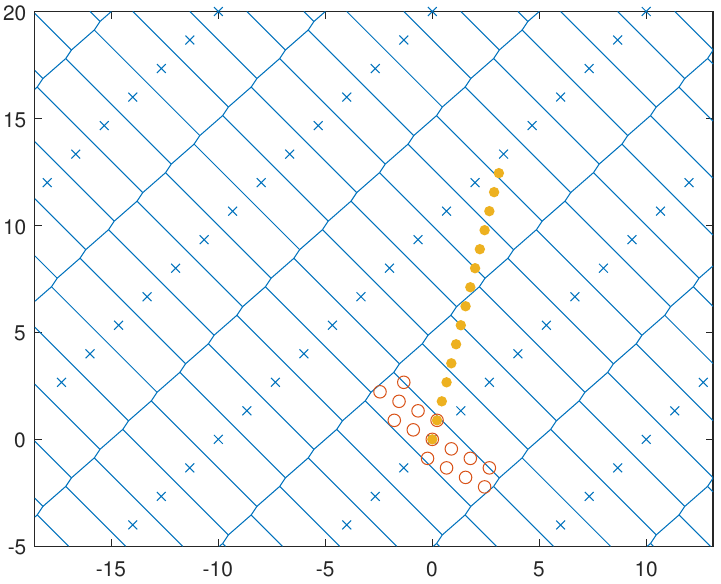}
  \caption{Rectangular encoding of cyclic lattice with isomorphism, $M=15$}
  \label{fig:iso}
\end{figure}

 Fixing the first row as in (\ref{eqn:example1}), solution to (\ref{eqn:isomorphismnd}) is $(r_1, r_2)=(-1, -2)$. Setting $M=15$, we could get an $n=2$ cyclic lattice possessing group isomorphism, as shown Fig. \ref{fig:iso}.

\begin{figure}[htbp]
  \centering
  \includegraphics[scale=0.5]{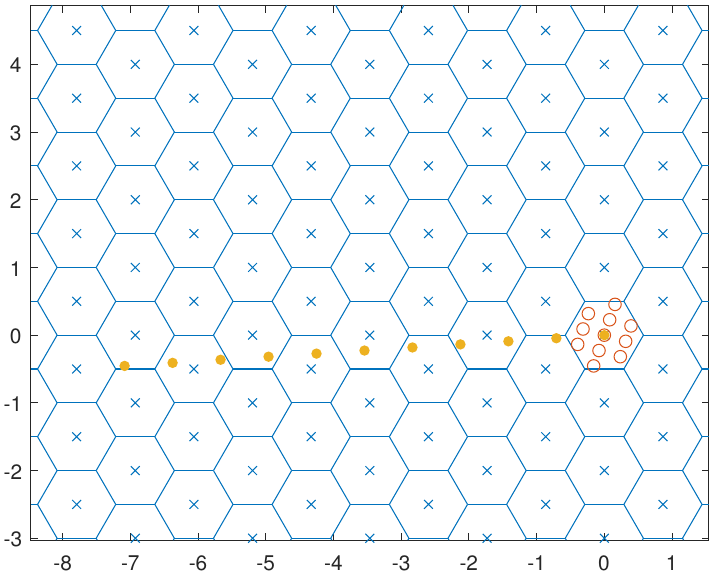}
  \caption{Rectangular encoding of cyclic lattice using $A_2$ shaping lattice, $M=11$}
  \label{fig:a2}
\end{figure}

 Since the $A_2$ lattice has the best shape in 2 dimensions, we also form a nested lattice code of arbitrary size $M$ using shaping lattice $A_2$ with generator matrix given by
\begin{equation}
    \label{eqn:isomorphic form-A2}
    \renewcommand\arraystretch{1.2}
    \*G_{A_2} = \begin{bmatrix}
 \frac{\sqrt{3}}{2} & 0\\
\frac{1}{2} & 1
\end{bmatrix}.
 \end{equation}
 Using the same $(a, b, r_1, r_2) = (4, 9, -1, -2)$ as in the last example, and setting $M=11$, we could obtain the isomorphic and cyclic nested lattice code as shown in Fig \ref{fig:a2}.


\subsection{$E_8$ lattice}
\begin{figure}[htbp]
  \centering
  \includegraphics[width=0.40\textwidth]{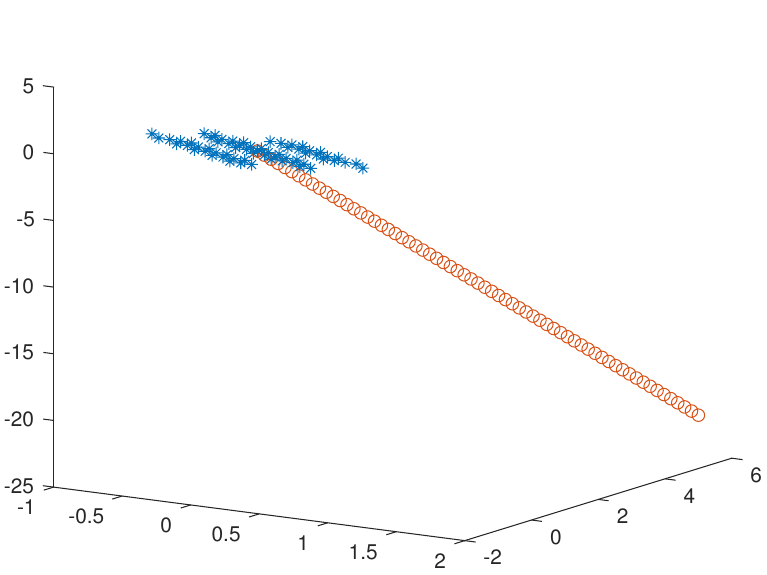}
  \caption{Rectangular encoding of cyclic lattice using $E_8$ shaping lattice, $M=64$}
  \label{fig:3de8}
\end{figure}

Consider the $E_8$ lattice as a shaping lattice and design $\*W$ with structure in (\ref{eqn:ndexist-fix3}) where $a=7, b=17, c=19$. One possible solution to 
(\ref{eqn:isomorphismnd-fix3}) is $r_6=95, r_7=65, r_8=92$ and $r_i =0, i \leq 5$. Setting $M=64$, we can obtain an isomorphic cyclic lattice code $\mathcal{C}$, where the coding lattice $\*G_c = \*G_s\*W^{-1}$. Since we use $E_8$ as shaping lattice, the shaping gain is unchanged, i.e., $0.65$ dB. $\mathcal{C}$ is in the subspace $\mathbb{R}^3$ which is shown in Fig.~\ref{fig:3de8}.

\section{Conclusion}
\label{sec:conclusion}
In this paper, we derived conditions for the existence of cyclic lattice codes and proposed two simple designs for general $n$ dimensional cyclic lattice codes. By solving linear Diophantine equations, these designs can be easily adapted to possess group isomorphism, which makes cyclic lattice codes a potential candidate for physical layer network coding. By fixing either the shaping lattice or coding lattice, we can design an isomorphic cyclic lattice code of arbitrary codebook size $M$. Cyclic lattice codes have shaping gain when a suitable shaping lattice is used. In this case, however, it might be time consuming to find a good coding lattice. The trade-off between coding gain and shaping gain is still unclear under this design. Nevertheless, it is possible to evaluate the performance with other constructions, e.g., \cite{5513739, tunali2015lattices, 7282518, 8387793,6952389} for physical layer network relaying techniques such as compute and forward \cite{6034734}, which remains to be our future work. 


\bibliographystyle{IEEEtran}
\bibliography{reference}

\begin{thebibliography}{10}
\providecommand{\url}[1]{#1}
\csname url@samestyle\endcsname
\providecommand{\newblock}{\relax}
\providecommand{\bibinfo}[2]{#2}
\providecommand{\BIBentrySTDinterwordspacing}{\spaceskip=0pt\relax}
\providecommand{\BIBentryALTinterwordstretchfactor}{4}
\providecommand{\BIBentryALTinterwordspacing}{\spaceskip=\fontdimen2\font plus
\BIBentryALTinterwordstretchfactor\fontdimen3\font minus \fontdimen4\font\relax}
\providecommand{\BIBforeignlanguage}[2]{{%
\expandafter\ifx\csname l@#1\endcsname\relax
\typeout{** WARNING: IEEEtran.bst: No hyphenation pattern has been}%
\typeout{** loaded for the language `#1'. Using the pattern for}%
\typeout{** the default language instead.}%
\else
\language=\csname l@#1\endcsname
\fi
#2}}
\providecommand{\BIBdecl}{\relax}
\BIBdecl

\bibitem{1056761}
J.~Conway and N.~Sloane, ``A fast encoding method for lattice codes and quantizers,'' \emph{IEEE Transactions on Information Theory}, vol.~29, no.~6, pp. 820--824, 1983.

\bibitem{8361838}
B.~M. Kurkoski, ``Encoding and indexing of lattice codes,'' \emph{IEEE Transactions on Information Theory}, vol.~64, no.~9, pp. 6320--6332, 2018.

\bibitem{6034734}
B.~Nazer and M.~Gastpar, ``Compute-and-forward: Harnessing interference through structured codes,'' \emph{IEEE Transactions on Information Theory}, vol.~57, no.~10, pp. 6463--6486, 2011.

\bibitem{zamir_nazer_kochman_bistritz_2014}
R.~Zamir, B.~Nazer, Y.~Kochman, and I.~Bistritz, \emph{Lattice Coding for Signals and Networks: A Structured Coding Approach to Quantization, Modulation and Multiuser Information Theory}.\hskip 1em plus 0.5em minus 0.4em\relax Cambridge University Press, 2014.

\bibitem{21245}
G.~Forney, ``Coset codes. {I}. introduction and geometrical classification,'' \emph{IEEE Transactions on Information Theory}, vol.~34, no.~5, pp. 1123--1151, 1988.

\bibitem{alonso1991notes}
J.~M. Alonso, T.~Brady, D.~Cooper, V.~Ferlini, M.~Lustig, M.~Mihalik, M.~Shapiro, and H.~Short, ``Notes on word hyperbolic groups,'' in \emph{Group theory from a geometrical viewpoint}, 1991.

\bibitem{Micciancio-FOCS02}
D.~Micciancio, ``Generalized compact knapsacks, cyclic lattices, and efficient one-way functions from worst-case complexity assumptions,'' in \emph{The 43rd Annual IEEE Symposium on Foundations of Computer Science, 2002. Proceedings}, 2002, pp. 356--365.

\bibitem{5513739}
C.~Feng, D.~Silva, and F.~R. Kschischang, ``An algebraic approach to physical-layer network coding,'' in \emph{2010 IEEE International Symposium on Information Theory}, 2010, pp. 1017--1021.

\bibitem{tunali2015lattices}
N.~E. Tunali, Y.-C. Huang, J.~J. Boutros, and K.~R. Narayanan, ``Lattices over {E}isenstein integers for compute-and-forward,'' \emph{IEEE Transactions on Information Theory}, vol.~61, no.~10, pp. 5306--5321, 2015.

\bibitem{7282518}
Y.-C. Huang, K.~R. Narayanan, and P.-C. Wang, ``Adaptive compute-and-forward with lattice codes over algebraic integers,'' in \emph{2015 IEEE International Symposium on Information Theory (ISIT)}, 2015, pp. 566--570.

\bibitem{8387793}
------, ``Lattices over algebraic integers with an application to compute-and-forward,'' \emph{IEEE Transactions on Information Theory}, vol.~64, no.~10, pp. 6863--6877, 2018.

\bibitem{6952389}
M.~A. Vázquez-Castro and F.~Oggier, ``Lattice network coding over euclidean domains,'' in \emph{2014 22nd European Signal Processing Conference (EUSIPCO)}, 2014, pp. 1148--1152.

\bibitem{sakzad2014phase}
A.~Sakzad, E.~Viterbo, J.~Boutros, and Y.~Hong, ``Phase precoded compute-and-forward with partial feedback,'' in \emph{2014 IEEE International Symposium on Information Theory}.\hskip 1em plus 0.5em minus 0.4em\relax IEEE, 2014, pp. 2117--2121.

\end{thebibliography}

\end{document}